\documentclass[10pt, conference]{IEEEtran}
\IEEEoverridecommandlockouts
\usepackage[switch]{lineno}
\usepackage{booktabs}
\usepackage{graphicx}
\usepackage{tcolorbox}
\usepackage{url}
\usepackage{enumitem}
\usepackage{cite}
\usepackage{hyperref}

\begin{document}

\title{An Exploratory Study on Build Issue Resolution Among Computer Science Students
\thanks{This work is supported in part by NSF projects 1736209 and 1846467, and the Google Cloud Research Credits program with the award GCP19980904. 

\textbf{Data Availability}: The dataset for this study is available on Zenodo (\href{https://doi.org/10.5281/zenodo.14885822}{https://doi.org/10.5281/zenodo.14885822}). Due to local IRB restrictions, some sensitive data may not be publicly accessible.}
}

\author{
\IEEEauthorblockN{Sunzhou Huang\IEEEauthorrefmark{1},
Na Meng\IEEEauthorrefmark{2},
Xueqing Liu\IEEEauthorrefmark{3},
Xiaoyin Wang\IEEEauthorrefmark{1}}

\IEEEauthorblockA{\IEEEauthorrefmark{1}\textit{Department of Computer Science, The University of Texas at San Antonio}, San Antonio, USA}
\IEEEauthorblockA{\IEEEauthorrefmark{2}\textit{Department of Computer Science, Virginia Tech}, Blacksburg, USA}
\IEEEauthorblockA{\IEEEauthorrefmark{3}\textit{Department of Computer Science, Stevens Institute of Technology}, Hoboken, USA}
\IEEEauthorblockA{sunzhou.huang@utsa.edu,
nm8247@vt.edu,
xliu127@stevens.edu,
xiaoyin.wang@utsa.edu}
}

\maketitle

\begin{abstract}
When Computer Science (CS) students try to use or extend open-source software (OSS) projects, they often encounter the common challenge of OSS failing to build on their local machines. Even though OSS often provides ready-to-build packages, subtle differences in local environment setups can lead to build issues, costing students tremendous time and effort in debugging. Despite the prevalence of build issues faced by CS students, there is a lack of studies exploring this topic. To investigate the build issues frequently encountered by CS students and explore methods to help them resolve these issues, we conducted a novel dual-phase study involving 330 build tasks among 55 CS students. Phase I characterized the build issues students faced, their resolution attempts, and the effectiveness of those attempts. Based on these findings, Phase II introduced an intervention method that emphasized key information (e.g., recommended programming language versions) to students. The study demonstrated the effectiveness of our intervention in improving build success rates. Our research will shed light on future directions in related areas, such as CS education on best practices for software builds and enhanced tool support to simplify the build process.
\end{abstract}


\section{Introduction}
It is highly beneficial to integrate open-source software (OSS) projects into a software engineering (SE) curriculum for three reasons. First, OSS exposes students to real-world software development practices. By inspecting the software evolution and reading developers' discussions on technical issues, students can improve their understanding of industrial challenges and become better prepared for careers in the software industry~\cite{Gokhale2012}. Second, OSS offers opportunities for students to gain hands-on  experience when they use and modify the software based on local needs and preferences~\cite{Kotwani2011}, collaborate with a global community of developers, and adopt project management tools like version control as well as issue tracking. Third, the educational usage of OSS bridges the gap between theoretical knowledge and practical application, enabling students to see how theoretical knowledge is applied in real life~\cite{Pinto2017} and to solve real problems with book knowledge.

While the integration of OSS projects into SE curriculum brings great opportunities, it also poses challenges to Computer Science (CS) education. One big challenge is that students often fail to build the provided OSS on their local machines. Build failures can be very detrimental to students' learning outcomes. This is because as students encountered build failures at the beginning of OSS-based class activities, they might spend tremendous time and effort trying to resolve those issues and neglect teachers' further instructions on designated tasks. As a result, they might lose the valuable opportunities to read, modify, or execute the same software as other students do and become unavailable to collaborate with other students. 

In the research area of CS education, educators and researchers proposed various ways of integrating OSS into CS curriculum~\cite{nascimento2015open,Pinto2017,Kotwani2011,Gokhale2012,choi2021open}, discussed the benefits and challenges brought by the OSS-integration into courses~\cite{Salerno_de2023}, and measured students' contributions to OSS~\cite{Fang_Endres_Zimmermann_Ford_Weimer_Leach_Huang_2023}. In particular, Salerno et al.~\cite{Salerno_de2023} conducted a large-scale study on the impact of OSS-based courses, revealing that 83\% of the students faced technical challenges while contributing to OSS projects. While only 2\% of students anticipated hurdles in setting up the local environment, a notable 9\% of students eventually reported facing such issues. This discrepancy suggests that students tend to underestimate the challenge of addressing build issues. None of the existing work characterizes the build issues CS students frequently encounter in OSS-based courses, let alone suggest intervention to help address those issues.

In the research area of build issues, some researchers conducted user studies to explore the influencing factors (e.g., tool usage) for software build results~\cite{Kwan_Schroter_Damian_2011,downs2012ambient,phillips2014understanding,kerzazi2014automated,hilton2016continuous,vassallo2020every}. Other researchers examined relevant artifacts (e.g., build logs and bug fixes) to characterize various aspects of software build, including developers' efforts in resolving build issues~\cite{mcintosh2011empirical,mcintosh2015large}, root causes for build failures~\cite{xia2014empirical,barrak2021builds,wu2020empirical}, and fixes~\cite{zhao2014empirical,lou2020understanding}. However, these studies are only vaguely related to the build issues faced by CS students for two reasons. First, prior work studied build issues from the perspective of developers or project managers who owned the software-to-build, rather than from the perspective of CS students who are new to the software-to-build. Second, prior studies were based on either self-retrospection or artifacts of experienced software practitioners, instead of based on the experience of CS students in class. Thus, many of the research findings by prior work are unlikely to find evidence in the build issues faced by students.

We conducted a related pilot study~\cite{huang2024build} to investigate the symptoms and causes of build issues experienced by non-contributors (e.g., users, learners, and potential contributors). The findings highlight specific build issues that are challenging to resolve and emphasize the need for further study to understand non-contributors' behavior.

To overcome the limitations of prior work, in this paper, we introduce an exploratory study to investigate the build issues faced and resolved by CS students. The study has two phases, which tracked the behaviors of 55 CS students as they undertook 330 OSS build tasks in an advanced SE course. In Phase I, each student of the course in 2022 was assigned six build tasks. Phase I collected and analyzed the 198 build results from 33 students to characterize the symptoms of build issues, the students' resolution attempts, and the effectiveness of those attempts. Phase II introduced an intervention to help students build software. Namely, the intervention emphasized providing additional key information sources, so that students can refer to them initially when completing build tasks. As with Phase I, Phase II assigned each student of the course in 2023 with six build tasks. By gathering and analyzing the 132 build results of 22 students, Phase II compared students' results with those from Phase I, to assess the intervention effectiveness.

This paper makes the following major contributions:
\begin{itemize}[leftmargin=*]
    \item We presented a comprehensive analysis of the build issues encountered by 55 CS students, offering insights into the common challenges faced during students' OSS builds.

    \item Our study revealed the resolution strategies frequently applied by students, although some of these strategies often led to build issues that are challenging to resolve.

    \item We introduced an intervention method (i.e., providing key information sources initially to bridge the knowledge gap), which demonstrated effectiveness in helping students better build OSS.

   \item Our study offers practical recommendations for various stakeholders in the OSS and CS education communities, aiming to enhance the overall experience of integrating OSS into the SE curriculum.
\end{itemize}

\section{Methodology}
Our exploratory study was designed and executed in two distinct phases to comprehensively understand build issue resolution among CS students. Each phase had a unique objective and contributed valuable insights to the overall study.

\subsection{Resolution Strategy Study}
The initial phase of our study is the resolution strategies study, focused on understanding the symptoms of unresolved build issues and analyzing strategies that CS students have employed to successfully address them. We collected data from 33 CS students across 4 OSS projects and 2 programming languages (PL) in the 2022 software testing class. We evaluated the effectiveness of these strategies and their limitations. In addition, we explored the impact of prior experience on the build success rate. This phase provided us with a deeper understanding of students' resolution strategies. RQ1-5 will be answered in this phase.

\begin{itemize}[leftmargin=*]

\item \textbf{RQ1}: As learners of OSS, are students able to accurately interpret the build issues that block build tasks?

\item \textbf{RQ2}: How often do build tasks end with unresolved issues?

\item \textbf{RQ3}: What are the common symptoms of unresolved build issues, and to what degree can they be mitigated?

\item \textbf{RQ4}: What strategies have students employed to successfully address these unresolved build issues, and are these strategies effective?

\item \textbf{RQ5}: Are there any prior experiences that may correlate with students’ ability to resolve build issues?

\end{itemize}

\subsection{Intervention Study} Phase II is the intervention study. The insights gathered from the preceding phase guided the development of intervention, which were aimed at enhancing the success rate of CS students on build tasks with proactive measures. The intervention was performed on 22 CS students registered for the same course in 2023. We assessed the impact of the intervention on the success rate and collected feedback to identify any potential mismatches. RQ6 will be answered in this phase.

\begin{itemize}[leftmargin=*]

\item \textbf{RQ6}: Given the resolution strategies employed by students, what proactive measures can we introduce to effectively enhance their performance in resolving build issues?

\end{itemize}

\subsection{Participants and Tasks}
Our participants were all CS students registered in the Software Testing course in 2022 ($n = 33$) and 2023 ($n = 22$). This course is a cross-listed elective for senior-year undergraduates and graduate students, with Software Engineering as a prerequisite. The course, taught by two of the authors at a U.S. university, focuses on software testing approaches, test planning, test case design, and build systems with CI/CD concepts. In our study, students only need to build the released OSS and do not need a background to modify the software.

Students were required to complete the study as one of their major course projects. We designed a three-stage task list to help participants become familiar with the study process:

\begin{enumerate}[leftmargin=*]

\item \textbf{Setup:} Participants were instructed to use the Google Cloud Platform (GCP) online console to set up a GCP project with \$50 education credits and enabled APIs for creating VMs. Then, custom images were shared with participants on GCP. Additionally, we obtained admin permission to access participants' GCP projects for further data collection.

\item \textbf{Warm-up:} Participants were instructed to use the warm-up custom image to create a VM instance. They then performed warm-up build activities in the VM, such as executing a ``Hello World'' build script. Participants learned to use the preinstalled script tool to log the VM state and then write a textual build report. They were also asked to complete a survey to provide feedback to researchers.

\item \textbf{Build OSS projects:} Participants were assigned OSS projects as their build task. They used preinstalled scripts and textual issue reports to report issues encountered while building and executing tests of the assigned project. At the end of this stage, participants completed a survey to share their feedback.

\end{enumerate}

All these OSS projects were in the release version and had been verified by the two authors who taught this course. The anticipated outcomes included compiling the assigned OSS projects and passing all tests using the commands specified in the build tasks. In addition to the build outcome, their grades would be evaluated based on their investment in completing the tasks through monitored VMs and issue reports. As the conductors of the study, our role was to help clarify the objectives of the build tasks. Our study protocol was assessed and approved by our local IRB. All students were notified that they could withdraw their data after the grading so that their data would not be included in our dataset for our study or any subsequent research activities.

Over a span of two years, our study involved a diverse set of students in a variety of tasks. Specifically, the participation was marked by 39 and 25 students in each phase, respectively. In 2022, we had a total of 33 participants who completed all tasks related to our resolution strategies study. The following year, in 2023, the number stood at 22 for the intervention study. Despite the fluctuation in participant numbers, the prerequisites and design of our course remained consistent.

\subsection{Data Collection}
\begin{figure}[tbp]
\centerline{
\includegraphics[width=\linewidth]{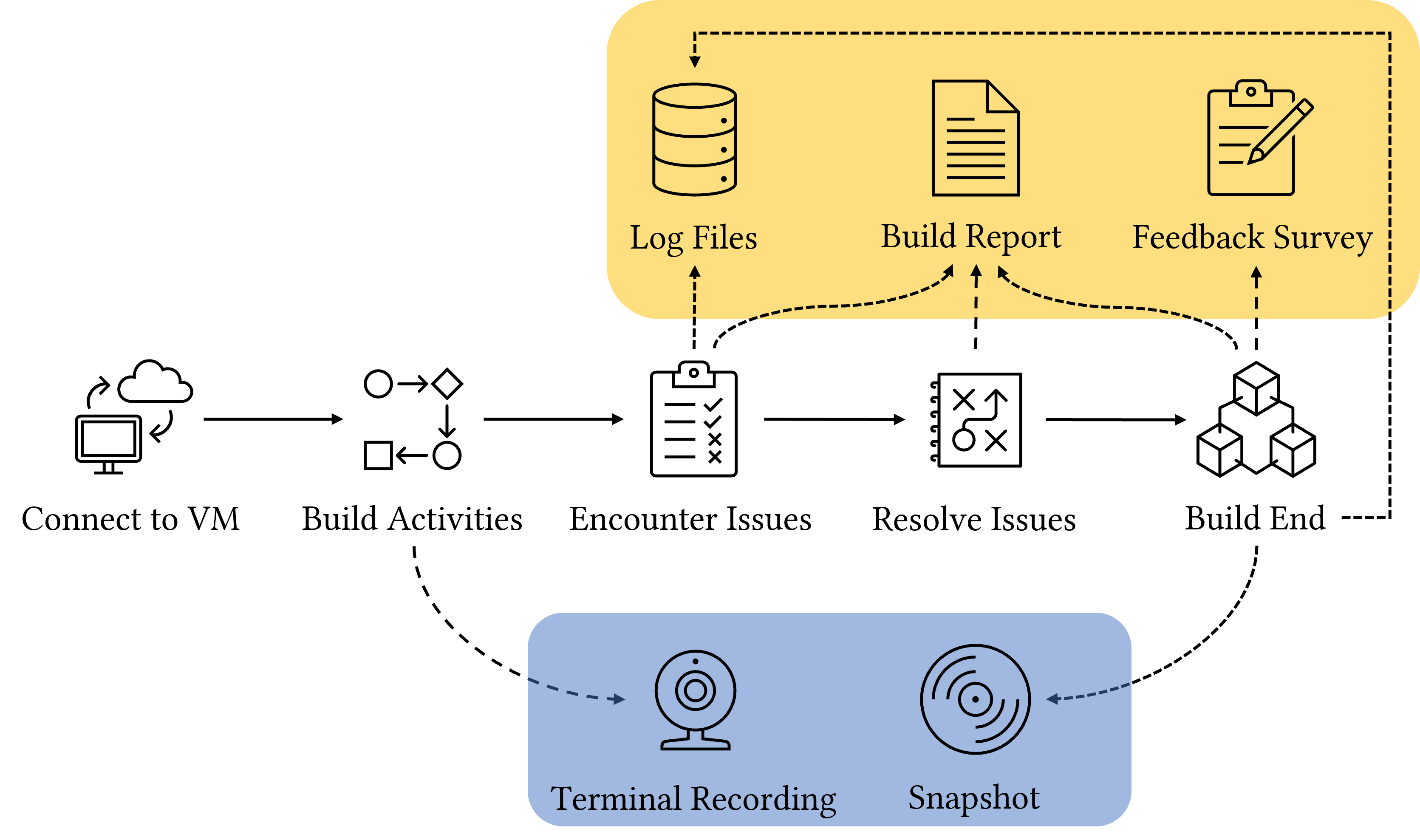}
}
\vspace{-.2cm}
\caption{Data Collection Framework for Build Tasks}
\vspace{-.4cm}
\label{fig:framework}
\end{figure}

We designed a framework that collects data throughout the entire build process, as illustrated in Figure~\ref{fig:framework}. While participants perform build activities on a custom VM and report build issues, all terminal activities are recorded. When a build task ends, the final state of the VM is preserved as a snapshot. The feedback survey is collected after all build tasks are completed. In Figure~\ref{fig:framework}, we denote active data collection with the upper yellow rounded rectangle and passive data collection with the lower blue rounded rectangle.

\begin{itemize}[leftmargin=*]
\item \textbf{Custom VM:}
In this study, we customized the official GCP OS images, which offer basic OS setups. We incorporated a custom script at OS startup to record terminal activities and provided a tool that logs the current state of the VM. These custom images ensure that all participants have consistent base build environments.
\item \textbf{VM state logging:}
After encountering build issues, participants log the current VM state using a script containing Linux commands to record command history, environment variables, and network information. This logging process, integral to the issue reporting, helps match logged data with reported issues. Additionally, participants log the final VM state upon completing the build tasks. All logs are saved in VMs for later retrieval, aiding in issue diagnosis and reproduction.
\item \textbf{Terminal recording:}
We opted for terminal recording as it captures all terminal activities, making the collected data easier to analyze. Using the \textit{script} command in Linux~\cite{script}, we integrated terminal recording into the VM's startup process. Once participants connect to the VM via SSH, the recording begins, capturing all screen outputs and keyboard inputs, including activities like file editing. While terminal recording lacks the ability to extract executable commands from inputs, it offers automated and comprehensive data collection without requiring manual initiation by participants. Despite this limitation, terminal recording facilitated playback and review of the entire build process, aiding in the analysis and reproduction of encountered build issues.
\item \textbf{Snapshots of VMs:}
In addition to terminal recording, another option for collecting VM data is taking snapshots. A snapshot captures the entire VM at a specific moment, including files, settings, and configurations. While useful for troubleshooting and reproducing issues, snapshots lack playback functionality and erase temporary environment variables. To address cost and practical concerns, we decided to utilize an automated script to capture snapshots only upon the build tasks reaching the final state, whether it be success or failure.
\item \textbf{Build issue report:}
Throughout the completion of all tasks, participants were directed to utilize the build issue report to document encountered issues during the build process, including inputs, outputs, and the solutions employed for resolution. Additionally, participants documented additional details of the build environment, such as program language versions and utilized tools. Upon completion of the build, participants summarized the entire process and provided their thoughts. The build issue report offers valuable insights into their decision-making and problem-solving strategies.

\item \textbf{Feedback Survey:}
After completing the build process and documenting the encountered issues and their corresponding solutions, we conducted a feedback survey to gather insights from participants. This survey focused on human factors affecting the build process, such as participants' familiarity with the PL, build system, and OS. Additionally, we collected self-reported build outcomes (success or failure). Insights collected from the comment section, which included obstacles and reflections, further aided in uncovering the challenges faced by participants during the build process.
\end{itemize}

After screening out participants who did not complete all tasks, we collected the following data for each phase:

\begin{enumerate}[leftmargin=*]
  \item \textbf{Resolution Strategies Study}: A total of 198 build results with snapshots containing terminal recordings and log files for the entire build process. In addition to 33 textual build issue reports with feedback surveys.
  \item \textbf{Intervention Study}: A total of 132 build results with snapshots containing terminal recordings and log files for the entire build process and 22 textual build issue reports along with feedback surveys.
\end{enumerate}

\subsection{Qualitative and Quantitative Data Analysis}
\subsubsection{Resolution Strategies Study}
\label{analysis2022}
To address \textbf{RQ1-2}, the first author manually verified the build outcomes reported by participants, cross-referencing them with terminal recordings, command histories, and environment information extracted from snapshots. If a build issue occurred that prevented participants from completing the build task, the error message was extracted, and it was labeled as \textit{unresolved}. Subsequently, all snapshots from the final state VM were converted into running VMs to reproduce the collected error messages. This was done using an automatic script, with necessary adjustments made for specific build issues. If there was an identified inconsistency in reporting, we refined the error message and resolution state. If a build issue could not be reproduced, the original error message was retained. The last author validated the verified build outcomes and the extracted error messages. If there were any conflicts, a discussion was initiated until a consensus was reached.
To address \textbf{RQ3-4}, two authors initially followed the open coding procedure~\cite{Seaman_1999} to label unresolved build issues with their root symptoms. These root symptoms were identified by examining the entire build process, which was based on terminal recordings and command histories. If any conflicts arose, further verification was performed on a running VM created from the snapshot, and a discussion was initiated until a consensus was reached. Next, based on the error messages, two authors counted the number of resolved build issues across all successful build tasks. We only include issues in successful build tasks to eliminate the uncertainty associated with chained failures, where a new failure might mask an old one. In addition, two authors identified common approaches that resolved these issues.
To address \textbf{RQ5}, we first mapped the answers from the 5-point Likert scale of familiarity questions to an ordinal scale, ranging from 1 to 5 (with 1 representing ``Not familiar at all'' and 5 representing ``Extremely familiar''). We then converted the build outcomes to a binary scale (1 for success, 0 for failure). Following this, we calculated the Spearman’s rank correlation coefficient to investigate any potential relationship between prior experiences and build outcomes. The significance of the relationship was interpreted with the $p$-value.

\subsubsection{Intervention Study}
To address \textbf{RQ6}, we performed the same verification process as in the resolution strategies study to obtain the verified build outcomes for the intervention study. We compared the build success rates before and after the intervention, and used the Chi-squared test ($\chi^2$) to determine if the two rates were significantly different. Furthermore, aside with a self-reported 5-point scale for helpfulness (ranging from 1-5, with an option for ``not applicable''), we investigated the barriers mentioned in textual feedback survey during the intervention study to identify the impact of intervention.


\begin{table}[tbp]
  \centering

  \caption{OSS projects in resolution strategies study}
  \vspace{-.2cm}
    \resizebox{0.8\linewidth}{!}{\begin{tabular}{lllll}
    \toprule
    Project & Version & PL    & Build Sys. & Repository \\
    \midrule
    \midrule
    Guava & 31.1  & Java  & Maven & google/guava \\
    Spring Boot & 2.6.12 & Java  & Gradle & spring-boot \\
    \midrule
    OpenCV & 4.6.0 & C++   & Cmake & opencv/opencv \\
    Bitcoin & 23    & C++   & Autotools & bitcoin/bitcoin \\
    \bottomrule
    \end{tabular}}%
    \vspace{-.4cm}
  \label{tab:project}%
\end{table}%

\section{Resolution Strategies Study}
\label{sec:results22}
To answer RQ 1-5, we conducted a study on resolution strategies in 2022 and analyzed 198 fully monitored build results from 33 participants.

In a background survey conducted in 2022, we found that the top four PLs our participants were familiar with are Java (100\%), Python (92.3\%), C/C++ (76.9\%), and JavaScript (64.1\%). Given that Java was the default PL for other course projects and considering that C++ requires extra compilation during the build process, we selected Java and C++ as our target PLs. We selected OSS projects from GitHub~\cite{github}. After filtering out projects that were tutorials, involved mixed PLs, or were not in English, we chose the most popular projects for two different build systems for each PL. This resulted in four OSS projects, as shown in Table~\ref{tab:project}. We believe that the build processes and challenges of the most popular OSS projects on GitHub can be largely generalized to other projects using the same PL and build system. As the study by Lou et al.~\cite{lou2020understanding} observed the impact of different OS on build issue resolution, we further selected three Linux distributions (Ubuntu 22.04 LTS, CentOS Stream 9, and openSUSE Leap 15.4) based on their popularity. We verified all projects on each OS to ensure they were ready to build and excluded some time-consuming test cases to match the computing resources available to participants. 

We randomly divided the participants into two groups. The group, which worked on the Guava-Java and OpenCV-C++ projects, ended up with 14 participants. The other group, which worked on the Spring Boot-Java and Bitcoin-C++ projects, concluded with 19 participants. All participants began with the Java projects and then proceeded to complete the C++ projects. The project guidelines for the Java and C++ projects are identical, with the exception that we included an estimated build time and machine types for the C++ projects. Despite having trained the participants on the configuration of VMs on GCP, we still observed that some participants working on the Spring Boot project complained about the long build time due to incorrectly assigning system resources to VMs. Considering that this situation could have a significant negative impact on the completion of the future tasks, we added a recommended configuration for system resources to address the insufficient system resource issues for the C++ projects.

\subsection{As learners of OSS, are students able to accurately interpret the build issues that block build tasks? (RQ1)}
\label{sec:RQ1}
We manually verified the 198 build outcomes reported by 33 participants using the approach mentioned in Section \ref{analysis2022}. We observed the following inconsistencies in reporting:
\begin{enumerate}[leftmargin=*]
    \item Misinterpretation of build results (31 out of 198): Participants misinterpreted the failed build results as successful. This was extensively observed in OpenCV (24, 57.1\%).
    \item Incorrect build command (1 out of 198): A wrong build command was used, resulting in a partial build outcome.
    \item Silent failure (2 out of 198): Some test cases were not executed due to a silent crash.
\end{enumerate}
In total, 34 out of 198 build outcomes were misinterpreted, resulting in a misinterpretation rate of 17.2\%. We believe this observation occurred due to two main reasons.
Firstly, the OpenCV project required non-standard environment variables to pass all the test cases. As participants noted that the failed test cases were only a small portion of the passed test cases, they tended to believe it was a successful build. This could pose a serious problem if the missing tests were to catch potential defects in the software.
Secondly, students lacked the background knowledge to fully interpret the build results since they had never seen a successful build of the new project before. They tended to scan the keywords in the build log, which led to the ignoring of silent failures.

\subsection{How often do build tasks end with unresolved issues? (RQ2)}
\label{sec:RQ2}
Table~\ref{tab:success} presents the success rate of build outcomes verified with the correction of inconsistent reporting. Spring Boot has the lowest success rate among the four PLs because it suffered significantly from insufficient system resources. OpenCV has the second-lowest success rate, mainly because most participants neglect a non-standard environment variable. openSUSE is not a popular OS in the documentation of our OSS projects, which leads to the lowest success rate of 30.3\%. The overall success rate is 45.5\%, indicating that more than half of the build tasks are blocked by unresolved issues.

\begin{table}[tbp]
  \centering
  \caption{Success rate of verified build outcomes}
  \vspace{-.2cm}
    \resizebox{0.9\linewidth}{!}{\begin{tabular}{l|rrr|r}
          & \multicolumn{1}{l}{Ubuntu} & \multicolumn{1}{l}{CentOS } & \multicolumn{1}{l|}{openSUSE } & \multicolumn{1}{l}{3 OS} \\
    \midrule
    Guava - 14 & 92.9\% & 100.0\% & 92.9\% & 95.2\% \\
    Spring Boot - 19 & 26.3\% & 15.8\% & 10.5\% & \textbf{17.5\%} \\
    Java - 33 & 54.5\% & 51.5\% & 45.5\% & 50.5\% \\
    \midrule
    OpenCV - 14 & 35.7\% & 35.7\% & 35.7\% & \textbf{35.7\%} \\
    Bitcoin - 19 & 68.4\% & 63.2\% & 0.0\% & 43.9\% \\
    C++ -33 & 54.5\% & 51.5\% & 15.2\% & 40.4\% \\
    \midrule
    Java \& C++ - 66 & 54.5\% & 51.5\% & \textbf{30.3\%} & \textbf{45.5\%} \\
    \end{tabular}}%
  \label{tab:success}%
  \vspace{-.4cm}
\end{table}%

\subsection{What are the common symptoms of unresolved build issues, and to what degree can they be mitigated? (RQ3)}
\label{sec:RQ3}
\begin{figure}[tbp]
\centerline{
\includegraphics[width=\linewidth]{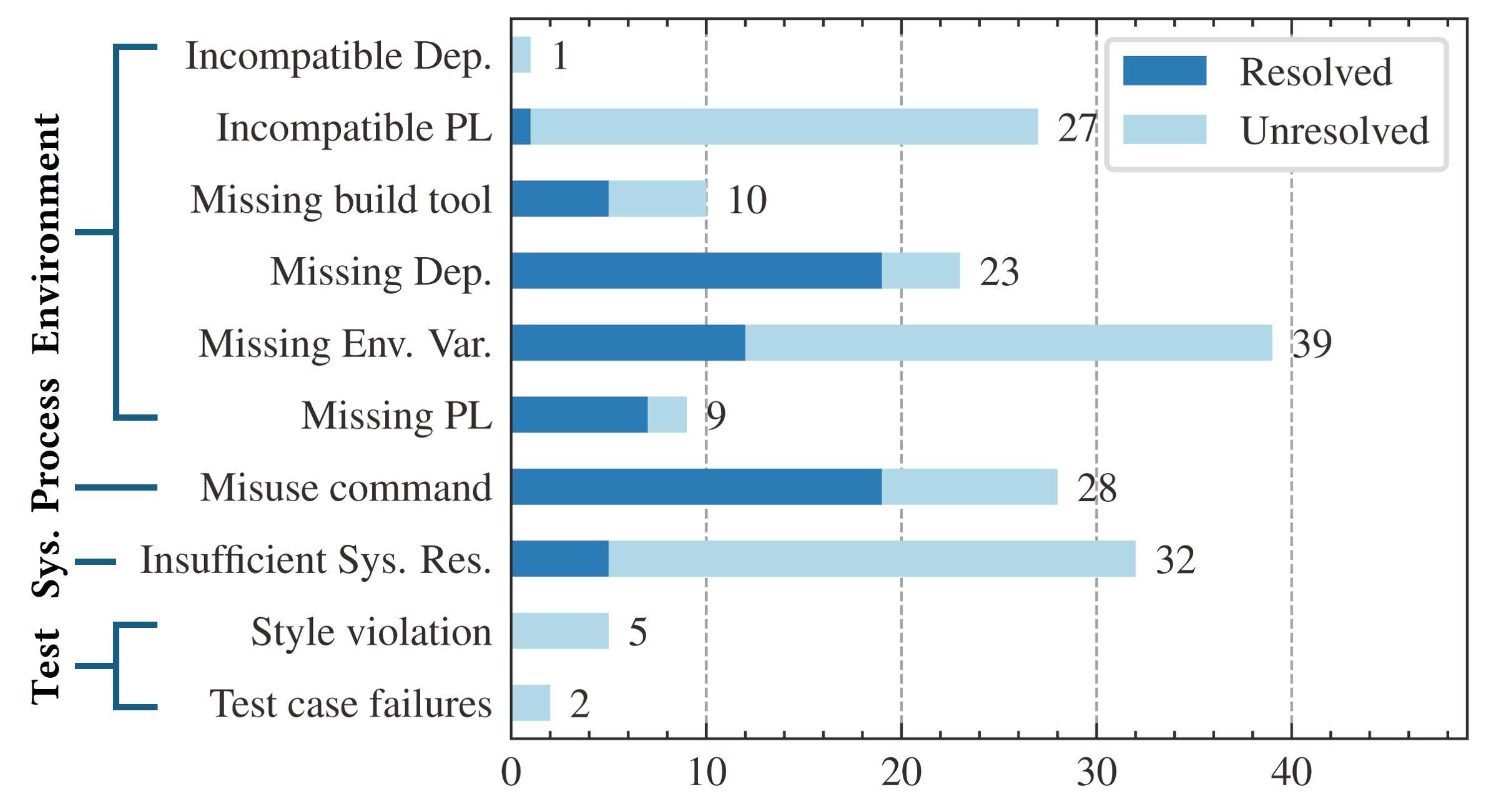}
}
\vspace{-.2cm}
\caption{Root symptoms from resolution strategies study}
\vspace{-.4cm}
\label{fig:symptoms22}
\end{figure}

In unresolved build issues, we identified 18 unique errors, as detailed in Table~\ref{tab:strategy}. Furthermore, we analyzed the resolved build issues that contain these 18 unique errors in successful build outcomes. In Table~\ref{tab:strategy}, ``\#UR'' represents the number of unresolved issues, while ``\#R'' denotes the number of resolved issues related to the 18 unique errors. Figure~\ref{fig:symptoms22} and Table~\ref{tab:strategy} illustrate these root symptoms, which can be categorized by Environment-Related (\textit{Incompatible} ... and \textit{Missing} ...), Process-Related (\textit{Misuse command}), System-Related (\textit{Insufficient system resources}), and Test-Related (\textit{Style violation} and \textit{Test case failure}).

The number of unresolved incompatible and missing issues indicates that environment-related issues constitute a significant portion of the failed build tasks. The build issues from the OpenCV project, which lacked a non-standard environment variable, contributed to the \textit{Missing environment variable}. Participants who built the Spring Boot project encountered a severe \textit{Insufficient system resources}. Many participants made mistakes by forgetting to configure the Makefile, leading to \textit{Misuse command}. 

We observed that the symptoms were not uniformly distributed. Consequently, students often encountered difficulties in resolving build issues caused by these recurring symptoms. Compared to the study by Lou et al.~\cite{lou2020understanding}, our study shows that students still frequently experience the same symptoms as developers. However, these are mostly limited to issues caused by a lack of installation, version incompatibility errors, and environment variable issues. We also observed additional symptoms not covered by previous studies that are related to students’ actions, such as \textit{Misuse command} and \textit{Insufficient system resources}.

\begin{table*}[htbp]
  \centering
  \caption{Errors, symptoms and resolutions}
    \resizebox{\linewidth}{!}{\begin{tabular}{lp{13em}p{40em}llr}
    \toprule
    Project & \multicolumn{1}{l}{Symptom} & \multicolumn{1}{l}{Error \& Description} & \#UR  & \#R   & \multicolumn{1}{l}{Resolution} \\
    \midrule
    Guava & \multicolumn{1}{l}{Insufficient Sys. Res.} & E1: Failed to execute goal org.apache.maven.plugins: maven-surefire-plugin:2.7.2:test (default-test) on project guava-testlib : There are test failures & 2     & 0     &  \\
    \midrule
    \multicolumn{1}{p{5.355em}}{Spring-Boot} & \multicolumn{1}{l}{Insufficient Sys. Res.} & E2: Gradle build daemon disappeared unexpectedly (it may have been killed or may have crashed) & 4     & 1     & \multicolumn{1}{l}{Retry with incremental compilation} \\
          & \multicolumn{1}{l}{} & E3: Build is slow & 7     & 0     &  \\
          & \multicolumn{1}{l}{} & E4: Build crash & 10    & 2     & \multicolumn{1}{l}{Retry with incremental compilation} \\
\cmidrule{2-6}          & Incompatible PL (Java) & E5: Execution failed for task ':spring-boot-project: spring-boot:compileKotlin' & 4     & 1     & \multicolumn{1}{l}{Use JDK instead of JRE} \\
          & \multicolumn{1}{l}{} & E6: Execution failed for task ':spring-boot-project: spring-boot:javadoc' & 11    & 0     &  \\
          & \multicolumn{1}{l}{} & E7: Execution failed for task ':buildSrc:test' & 1     & 0     &  \\
          & \multicolumn{1}{l}{} & E8: Execution failed for task ':spring-boot-project: spring-boot:test'. 4 test failures & 1     & 0     &  \\
\cmidrule{2-6}          & Missing PL (Java) & E9: ERROR: JAVA\_HOME is not set and no 'java' command could be found in your PATH & 2     & 7     & \multicolumn{1}{l}{Install Java} \\
\cmidrule{2-6}          & \multicolumn{1}{l}{Style violation} & E10: Execution failed for task ':checkstyleNohttp' & 5     & 0     &  \\
\cmidrule{2-6}          & Test case failures (Flaky) & E11: Execution failed for task ':spring-boot-project: spring-boot:test'. flaky test cases & 2     & 0     &  \\
    \midrule
    \multicolumn{1}{p{5.355em}}{OpenCV} & \multicolumn{1}{l}{Missing Env. Var.} & E12: 7 FAILED TESTS & 27    & 12    & \multicolumn{1}{l}{Set up OPENCV\_TEST\_DATA\_PATH} \\
    \midrule
    \multicolumn{1}{p{5.355em}}{Bitcoin} & \multicolumn{1}{l}{Insufficient Sys. Res.} & E3: Build is slow & 3     & 1     & \multicolumn{1}{l}{Wait patiently} \\
          & \multicolumn{1}{l}{} & E4: Build crash & 1     & 1     & \multicolumn{1}{l}{Build in background } \\
\cmidrule{2-6}          & Missing build tool (Make) & E13: make: command not found & 5     & 5     & \multicolumn{1}{l}{Install Make} \\
\cmidrule{2-6}          & Misuse command (no configure) & E14: make: *** No rule to make target 'check'.  Stop & 9     & 19    & \multicolumn{1}{l}{Configure Makefile} \\
\cmidrule{2-6}          & \multicolumn{1}{l}{Missing Dep.} & E15: Libtool library used but 'LIBTOOL' is undefined & 1     & 11    & \multicolumn{1}{l}{Install Libtool library } \\
          & \multicolumn{1}{l}{} & E16: configure: error: C++ compiler cannot create executables & 3     & 8     & \multicolumn{1}{l}{Install build-essential} \\
\cmidrule{2-6}          & \multicolumn{1}{l}{Incompatible Dep.} & E17: Segmentation fault & 1     & 0     &  \\
\cmidrule{2-6}          & Incompatible PL (G++) & E18: configure: error: cannot figure out how to use std::filesystem & 9     & 0     &  \\
    \bottomrule
    \end{tabular}}%
  \label{tab:strategy}%
  \vspace{-.4cm}
\end{table*}%

\subsection{What strategies have students employed to successfully address these unresolved build issues, and are these strategies effective? (RQ4)}
\label{sec:RQ4}
Based on the resolved build issues related to the unresolved build issues (observed in RQ3),  we studied the resolutions employed by students and summarized them into two strategies:  \textit{Brute Force} and \textit{Ad Hoc}, which seem not very effective as the overall resolution rate ($\frac{\sum \#R}{\sum (\#UR+\#R)}$) is  38.6\%.

\textbf{Brute Force:} This strategy has been applied to most instances of the \textit{Insufficient system resources} build issue. Benefiting from the incremental compilation feature of Gradle in the Spring Boot project, a few participants who encountered system-related build issues were able to overcome low resource issues. However, for most participants who suffered low resource issues, this strategy tended to fail. Our observations indicate that most participants were overly focused on this strategy, and the failure of each retry led to a tremendous sense of frustration.

\textbf{Ad Hoc:} Most participants applied the \textit{Ad Hoc} strategy, which tended to be effective for \textit{missing}-related symptoms. Participants simply needed to install the component that was reported as missing in the error message. However, for non-obvious symptoms, such as the \textit{missing environment variable} symptom, it was easy for participants to overlook potential failures, not to mention the incompatibility issues. This strategy may further increase the incidence rate of encountering \textit{trap issues}.

We further observed that some build issues, referred to as \textbf{trap issues}, can be the root cause of the low effectiveness of students’ strategies. Based on the two common strategies, we identified several issues that tend to lead to a failed build for students. For instance, the issues with E5 can be resolved by changing the Java Runtime Environment (JRE) to the Java Development Kit (JDK), as indicated in the detailed error message ``Make sure Kotlin compilation is running on a JDK, not JRE.''. However, changing from JRE to JDK is more complex than installing JDK initially. Even with this explicit error message, other participants were still unable to resolve the issue. Another significant example is the issue associated with E10. In this scenario, participants downloaded the JDK package directly into the root directory of the Spring Boot project. This action led the Checkstyle tool to inadvertently verify the JDK package, which ultimately resulted in the test crashing.

In essence, \textit{trap issues} are build issues that contradict students’ intuition. Consequently, build issues with a low resolution rate are considered \textit{trap issues}. Predicting \textit{trap issues} without sufficient information can be challenging. In this study, the incompatible issues and style violation issues can be identified as \textit{trap issues} since most of them have a low resolution rate. The official documents of OSS usually don’t contain recommended system resource information for builds. Therefore, the \textit{insufficient system resources} build issue can be identified as a trap issue for students. Neither \textit{brute force} nor \textit{ad hoc} strategies can effectively resolve these \textit{trap issues}, turning students’ build activities into a game of chance.

\subsection{Are there any prior experiences that may correlate with students’ ability to resolve build issues? (RQ5)}
\label{sec:RQ5}

\begin{figure}[bp]
\vspace{-.4cm}
\centerline{
\includegraphics[width=.9\linewidth]{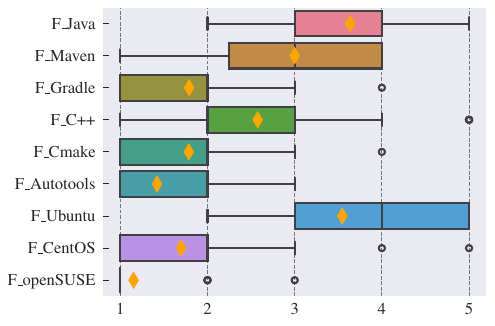}
}
\vspace{-.4cm}
\caption{Distribution of prior experiences}
\label{fig:farmiliarity}
\end{figure}

\begin{figure}[tbp]
\centerline{
\includegraphics[width=\linewidth]{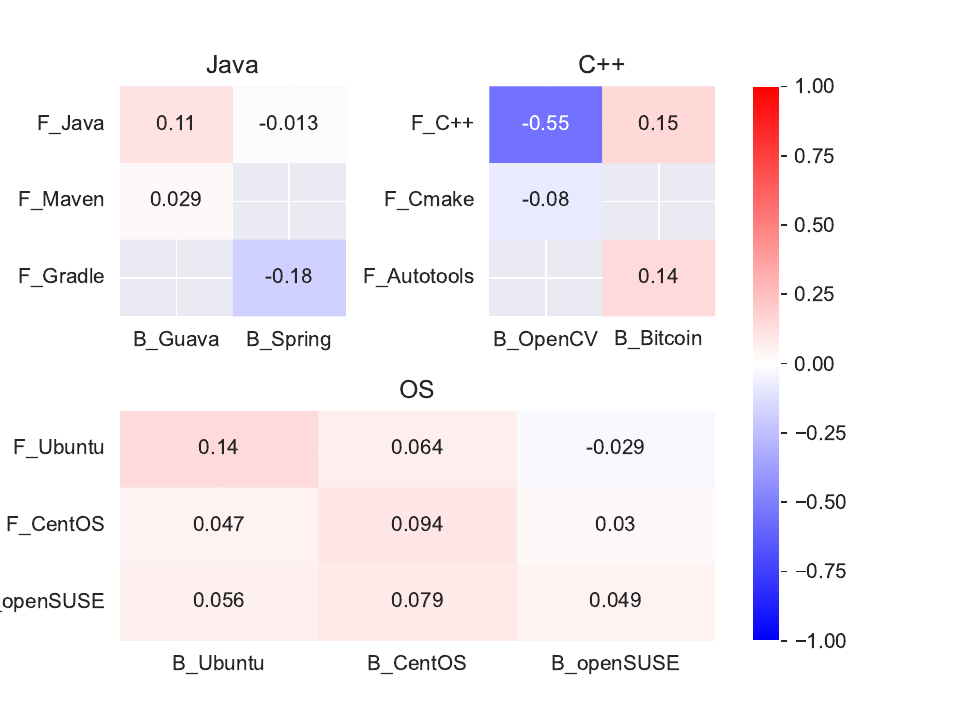}
}
\vspace{-.4cm}
\caption{Spearman’s rank correlation of prior experiences on build success}
\label{fig:corr}
\vspace{-.4cm}
\end{figure}

Figure~\ref{fig:farmiliarity} presents the distribution of familiarity, ranging from 1 to 5 (with 1 representing ``Not familiar at all'' and 5 representing ``Extremely familiar'').
Most participants are more familiar with Java and Ubuntu than with other PLs and OSs. We calculated the Spearman’s rank correlation coefficient between each type of familiarity and the build outcomes.
Figure~\ref{fig:corr} presents the coefficients, where ``F\_'' denotes familiarity and ``B\_'' denotes build outcome.

Most prior experiences have a very weak correlation with build outcomes. However, the prior experience with C++ and the build outcome of OpenCV have a moderate negative correlation (-0.55, $p$-value$<$0.001). We further investigated the collected data and confirm the trend: participants who are more familiar with C++ tend to fail more often in the OpenCV project (Success rates: Not familiar at all=66.7\%, Slightly familiar=50\%, Moderately familiar=0\%, Very familiar=0\%).
Since the majority of participants misreport their OpenCV build outcomes, a potential reason could be that participants more familiar with C++ tend to overlook details in the official documents, which leads to failures with E12.

The rest of the prior experiences, such as PLs, build systems, and OSs, do not show significant correlation. As correlation does not imply causation, the impact of prior experiences on students’ ability to resolve build issues may still need further exploration.

\section{Intervention Study}
\label{sec:results23}
In our resolution strategies study, we observed that participants frequently fell into \textit{trap issues}. We believe this is the main reason why students were unable to resolve the build issues. We hypothesize that preventing these \textit{trap issues} could be a more effective strategy than the common strategies currently employed by students. To test this hypothesis, we conducted an intervention study in 2023, where we added additional references to the project guidelines for participants. To be practical and follow most scenarios of using OSS in class, we allow students to use any external references they can find in both phases. The official documents were already provided in the first phase. As previously noted, the previous group that worked on the Spring Boot (Java) and Bitcoin (C++) projects encountered a wider variety of symptoms. Therefore, for 2023 participants, we selected Spring Boot and Bitcoin projects with the same release versions. We kept the rest of the study setup as identical as possible.

\subsection{Given the resolution strategies employed by students, what proactive measures can we introduce to effectively enhance their performance in resolving build issues? (RQ6)}
\label{sec:RQ6}
The additional references provided for the intervention are as follows:

\begin{itemize}[leftmargin=*]
    \item \textbf{Official guide links with ``how to build''}: in the 2022 study, we only provided links to general official guides. However, in the 2023 study, we improved this by providing links specifically to the ``how to build'' sections.
    \item \textbf{Example output for successful build results}: the screenshots demonstrate successful build outcomes. We also clarified the unrelated error message that appears in the CI reporting for the Spring Boot project.
    \item \textbf{Recommended PL versions}: we extract recommended PL versions from official documents and list them explicitly.
    \item \textbf{Dependencies}: information about how dependencies are required to be installed in different build systems, with links to the dependencies sections in official documents.
    \item \textbf{JDK vs. JRE}: we provide a link to the difference between JDK and JRE in Java, and remind that the package name might be different in Linux distributions.
    \item \textbf{CheckStyle}: background knowledge of the Checkstyle plugin and what it will do to a Spring Boot project.
    \item \textbf{Recommended system resource configuration for VMs}: as we mentioned in our study on resolution strategies, the information was initially provided only for C++ projects. In this phase, we have also extended it to Java projects. configuration
\end{itemize}

\begin{table}[tbp]
  \centering
  \caption{Build Outcomes with intervention}
    \resizebox{\linewidth}{!}{\begin{tabular}{lllll}
    \toprule
          & \multicolumn{2}{c}{Spring Boot - Java} & \multicolumn{2}{c}{Bitcoin - C++} \\
          & $\chi^2=64.2$ & $p<.001$ & $\chi^2=17.6$ & $p<.001$ \\
\cmidrule(lr){2-3}\cmidrule(lr){4-5}          & Success & Failure & Success & Failure \\
    \midrule
    Baseline - 57 & 10 (\textbf{17.5\%}) & 47 (82.5\%) & 25 (\textbf{43.9\%}) & 32 (56.1\%) \\
    Intervention - 66 & 60 (\textbf{91\%}) & 6 (9\%) & 54 (\textbf{82\%}) & 12 (18\%) \\
    \bottomrule
    \end{tabular}}%
    \vspace{-.4cm}
  \label{tab:interverntion}%
\end{table}%

As shown in Table~\ref{tab:interverntion}, the success rates for both projects were improved by intervention. The success rates of Spring Boot and Bitcoin improved from 17.5\% to 91\% and 43.9\% to 82\%, respectively. These improvements are statistically significant (both $p$-values of the Chi-squared test are less than 0.001).

\begin{figure}[bp]
\vspace{-.6cm}
\centerline{
\includegraphics[width=.9\linewidth]{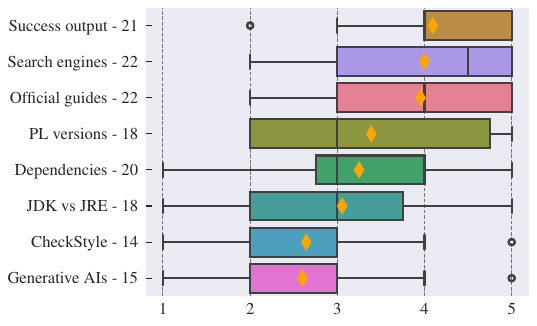}
}
\vspace{-.4cm}
\caption{Helpfulness rating for intervention (diamond markers indicates means)}

\label{fig:help}
\end{figure}

Figure~\ref{fig:help} presents the box plot of helpfulness rating from participants about intervention, ranging from 1 to 5, with an option for ``not applicable''. The example output (success output) outperforms other references. The search engines and official guides also have similar means. The references to PL versions and dependencies are helpful, and students also explicitly mentioned their usefulness in the barriers section of the feedback surveys. References to ``JDK vs. JRE'' and ``CheckStyle'', which are not found in official documents, rated to be less helpful than others. However, our observations from the feedback surveys indicate that participants did benefit from these external references. We did not observe any Checkstyle-related issues among the unresolved issues, and two participants explicitly stated that the barrier for them is the difference between JRE and JDK.

For the Spring Boot project, all six unresolved issues are incompatibility issues, which are the same as the \textit{trap issues} identified in our 2022 study. One of the six unresolved issues was actually resolved after rebooting the VM. The official OSS documents usually don’t include recommended system resource information for builds. Since we provided the recommended system resource configuration in this phase, none of the participants reported symptoms of insufficient system resources. The intervention on this OSS project was observed to reduce both \textit{trap issues} and other build issues. However, self-reported helpfulness does not necessarily equate to actual effectiveness for all students.

For the Bitcoin project, we observed that 1 of the 12 issues was due to low system resource. Also, a \textit{Misuse command (no configure)} issue was observed in the 12 unresolved issues. 10 of the 12 unresolved issues were related to incompatibility, 9 of which occurred in the openSUSE OS. In the openSUSE, for the Bitcoin project, the occurrence rate of \textit{trap issues} (incompatible) reduced from 52.6\% (10/19) to 40.9\% (9/22), and other build issues were fully resolved after intervention. As the recommended system resource information was provided to students in all phases of the Bitcoin project, the intervention was observed to reduce both trap issues and other build issues, regardless of the system resource limits.

When looking at each student individually, all the references provided are beneficial to them. This is because every reference has been rated with a highest helpfulness score of either 4 or 5, as shown in Figure~\ref{fig:help}. Most of these helpful references can be easily obtained from an OSS project and used as proactive measures, such as well-organized official guides, recommended PL versions, and dependencies. The success output is difficult to extract from documents but could be accessed from the CI system or previous build logs. Identifying and presenting project-specific knowledge, such as ``JDK vs. JRE'' and ``CheckStyle'', to students as a proactive measure poses a challenge. Despite often being overlooked by document contributors, these project-specific knowledge may still play a critical role in preventing trap issues for students.

\section{LESSONS LEARNED}
\label{sec:lesson}
We categorize our lessons learned into five subsections based on the various stakeholders in the OSS and CS education communities.

\subsection{Educators}

As a general finding from our study, students often face build issues when they try to build real-world OSS projects. Therefore, educators should be aware of such challenges whenever they adopt real-world OSS projects in their SE course projects.

As students are often unfamiliar with the OSS build scenario and lack project-specific experience, misinterpretations of build issues do occur. Section~\ref{sec:RQ1} reveals a 17.2\% misinterpretation rate in our exploratory study, which is not negligible. A list of commonly faced build error messages and their solutions could help students and prevent them from struggling without understanding the root causes. Furthermore, if educators rely on students’ self-reported evaluations to design or adjust SE courses related to OSS builds, additional clarification or deeper investigation should be performed to address this mismatch.

The educational use of OSS in CS courses allows students to work with real-world projects, making it an important aspect of the SE curriculum. However, as Section~\ref{sec:RQ2} shows, build issues can pose significant barriers to the successful integration of OSS in course projects. Students’ strategies for resolving build issues can be vulnerable, and they often fall into \textit{trap issues}, which require significant time and effort to resolve. Therefore, educators should select reliable OSS projects, identify \textit{trap issues} early, and implement proactive measures to mitigate the difficulties associated with the OSS projects build.

In the meantime, Section~\ref{sec:RQ6} shows that there might not be a silver bullet for proactive measures, so continuing to develop these measures could be a more viable approach. Additionally, Section~\ref{sec:RQ5} reveals that prior experiences may not apply to builds. If OSS builds tend to be project-specific, gathering project-specific experiences from student feedback or key information from extensive OSS documents could provide better assistance to students than relying solely on their expertise and background.

\subsection{Researchers}
Build issues often arise while students work on OSS projects, presenting both a challenge and a research opportunity. Section~\ref{sec:RQ3} shows that these build issues are not only similar to a subset of developers’ build issues but can also be unique from a student’s perspective, warranting deeper exploration across a wider range of OSS subjects. Identifying \textit{trap issues} is essential for addressing build issues. However, in cases where students have limited information, such as when they are working on an OSS project newly integrated into a CS course, it may be difficult for them to identify these issues. Future research could focus on developing more precise and effective methods for identifying \textit{trap issues} so that students can avoid them and educators can prepare guidance on how to bypass them. Additionally, exploring the impact of prior experiences on build processes, as indicated by Section~\ref{sec:RQ5}, could be another area for future research. Collecting a more refined dataset and involving a larger group of students may yield different results. Proactive measures have been proven effective in reducing build issues, but not all measures can be easily created without project-specific knowledge or key information. Therefore, research focused on extracting information for proactive measures is worth exploring.

According to Section~\ref{sec:RQ1}, inconsistencies in students’ self-reported build issues cannot be ignored. To avoid these threats, it is more effective to monitor the process in real-time rather than reproduce it later. While verifying the reported data can be time-consuming, it may be necessary.

\subsection{Students}

As students engage in OSS build activities, they take on the role of learners. To avoid common pitfalls, several pieces of advice can be beneficial. Previous experiences with PLs or build systems may not be relevant to the current OSS project. Therefore, it is important to rely on the key information from OSS documentation. Students should be encouraged to try different strategies, such as resetting the build environments, even if they are not comfortable with them. Intuitive strategies might lead to deeper issues. Sometimes, students need to think outside the box, as some issues may be caused by external resources or even flaky tests. Recognizing their own inexperience is crucial, which includes accepting the possibility of misinterpretation and a lack of project-specific knowledge. This means that students may need additional support while building OSS, even if they appear confident in related areas. If possible, using a popular OS can improve the build success rate.

Students' experiences of build failure can be valuable if reported to educators or OSS contributors, as this feedback can help create proactive measures for future students.

\subsection{OSS contributors}

OSS contributors should recognize the increasingly significant role that OSS plays in CS education, which also facilitates students' participation in OSS projects. Given that students may not have the required knowledge, particularly about OSS projects builds, mutual understanding of different roles within the OSS community is critical. Adding some external references, even if they pertain to basic background knowledge, can be helpful for learners. Not only are project-specific experiences valuable for other roles, but well-organized documentation and public build information are also crucial. Our study finds that error logs from the compiler or build systems are often too general to be useful and can sometimes be misleading. OSS contributors should consider supplementing them with more project-specific information to avoid misinterpretations. Build error messages are also sometimes hidden due to the complicated build process (e.g., the compilation process was called by a script but the log was not preserved or passed to the caller process). So build scripts need to be designed and tested more carefully to explicitly show error messages.

The project-specific experiences of OSS contributors can lead to the development of proactive measures that ultimately benefit OSS itself. Understanding students’ needs can help them become potential contributors in the future.

\subsection{Generative AI}
As generative AIs became increasingly popular in 2023, we noted that using AI services may impact our intervention study compared with common external references found by search engines. We did not limit students’ use of any external references in both phases to be practical. Figure~\ref{fig:help} shows that students rated generative AIs as the least helpful (lowest mean). To understand the reason behind this observation, we further case-studied the AI services by querying some error messages on ChatGPT 3.5 and 4~\cite{openai2024chatgpt} in February 2024. The results of the case study align with the students’ feedback. For example, we tested the error ``Execution failed for task :spring-boot-project:spring-boot-tools:spring-boot-antlib:javadoc.'' ChatGPT 3.5 gave 10 solutions, but only one of them mentioned the real root cause (JDK compatibility). ChatGPT 4 gave 3 unrelated solutions. The reason could be that AI services need key information to provide accurate suggestions in build issue resolution scenarios, but students may lack the experience to extract such information as input to AI services. Therefore, the answers from AIs were often overly general for students to find them helpful, because (1) so many different solutions may make students confused about which direction they should move to, and (2) the solutions are high level and do not include concrete actionable steps for the students to follow. In this OSS build scenario, future work may include adding more context to generative AIs, such as providing additional references to generative AIs.
\section{Threats to Validity}
\subsection{Internal Validity}
The major threat to the internal validity of our study is the reliance on self-reported data and manual verification processes. There may be errors in the process of data recording and analysis. To reduce this threat, we introduced a conflict resolution procedure and followed the open coding~\cite{Seaman_1999}. In addition, we leverage multiple information channels (e.g., reports, screenshots, command logs, and terminal recordings) for data recording and cross-validate the data among these channels.

Our study involving students who are one year apart can introduce a threat to internal validity. To mitigate this threat, we ensure that the curriculum and teaching methods remain consistent across the years to minimize differences in educational content. We use the same version of OSS projects and task setup and employ quantitative methods to analyze the overall build success rate. By collecting feedback from students, we aim to identify and address any potential issues that could affect the validity of our findings.

To ensure practicality, we permitted students to use external references, such as search engines, which may introduce threats to our study. Additionally, we did not track the usage of system resources, and providing a recommended system resource configuration to reduce the impact of build issues on low system resources could potentially weaken our findings. To mitigate these threats, we diligently analyze the collected data and investigate the details of each build issue in conjunction with our findings.

\subsection{External Validity}
The major threat to the external validity of our study is that our findings may not be generalizable to other students and OSS projects. To reduce this threat, we use popular OSS on GitHub~\cite{github} in different PLs, different build systems, and three OS variants to increase their representativeness. Given the exploratory nature of our study, our findings may not be generalizable to other institutions and courses that have different characteristics, such as varying enrollment capacities and diverse student demographics. Future research should consider these factors to enhance the applicability of the results across different educational settings.

\section{Related Work}
\label{sec:related}

Our study is related to existing research efforts on the educational use of OSS projects, user studies of software build, empirical studies of build issues from the developer's perspective, and studies of novice programmers. We will discuss related works in each category in the following subsections.

\textbf{Educational Use of OSS Projects in CS Courses}
The integration of open-source software (OSS) projects into computer science (CS) education has been widely explored. 
Nascimento et al.~\cite{nascimento2015open} reviewed how OSS projects facilitate software engineering (SE) learning. 
Choi et al.~\cite{choi2021open} proposed models for integrating OSS into lower-level CS courses. 
Fang et al.~\cite{Fang_Endres_Zimmermann_Ford_Weimer_Leach_Huang_2023} compared student contributions to general OSS projects and those aimed at social good, offering insights into educational interventions. 
Salerno et al.~\cite{Salerno_de2023} examined the impact of OSS development courses on students' self-efficacy and the barriers they face. 
These studies collectively underscore the value of OSS projects in CS education, highlighting both the challenges and the potential for enhancing student learning and engagement. Unlike previous studies, our research specifically focuses on the build challenges encountered during the educational use of OSS projects in CS courses.

\textbf{User Studies of Software Build.} The most relevant research to ours involves user studies on software build tasks. 
Kwan et al.~\cite{kwan2011does} examined IBM developers to see if team composition and coordination affect build success. 
Dawns et al.~\cite{downs2012ambient} evaluated a build management tool's impact on handling build failures.
Philips et al.~\cite{phillips2014understanding} studied Microsoft build teams, finding social challenges were significant.
Kerzazi et al.~\cite{kerzazi2014automated} analyzed 3,214 builds, noting a 17.9\% failure rate and significant time costs.
Hilton et al.~\cite{hilton2016continuous} interviewed developers about continuous integration (CI) processes.
Vassallo et al.~\cite{vassallo2020every} assessed a tool for summarizing build failures. 
Due to the challenges of performing such user studies, we can also see that there are not many existing studies, and all the studies have relatively small scales. 
Different from our research, all these studies are from the perspective of project managers or senior developers instead of CS students (OSS learners). 

\textbf{Empirical Studies on Build Issue}
Besides user studies, there have been many empirical studies on software build history and failures. McIntosh et al.~\cite{mcintosh2011empirical} studied version histories to estimate the effort required to revise and repair build scripts. Later, they correlated this effort with the type of build systems used~\cite{mcintosh2015large}. Xia et al.~\cite{xia2014empirical} summarized the characteristics of bugs in build systems. Zhao et al.~\cite{zhao2014empirical} found that build failures, though typically of lower severity, take more time to fix. Xia and Li~\cite{xia2014empirical} investigated the feasibility of predicting build failures using the TravisTorrent dataset~\cite{beller2017travistorrent}. Shridhar et al.~\cite{shridhar2014qualitative} qualitatively analyzed changes in build scripts. Barrak et al.~\cite{barrak2021builds} studied the correlation between build failures and code smells. Lou et al.~\cite{lou2020understanding} identified patterns in build fixes. Zolfagharinia et al.~\cite{zolfagharinia2017not} found different distributions of build failures across environments. Licker and Rice~\cite{licker2019detecting} used mutation testing to detect incorrect rules in build scripts. Wu et al.~\cite{wu2020empirical} studied build failures in Docker environments. 
These existing studies focus on build failure logs and build failure fixes in the commit history. In contrast, our study monitors and analyzes the whole process of CS students finishing multiple building tasks and records all the system environment factors that affect the build process failures.

\textbf{Studies of Novice Programmers}
Another research area related to our research is user studies of novice programmers. 
Lahtinen et al.~\cite{lahtinen2005study} surveyed over 500 students and teachers in computer-related majors to identify challenges faced by novice developers. Warner et al.~\cite{warner2017codepilot} evaluated CodePilot with eight novice developers to assess its usability and educational benefits. Marques et al.~\cite{marques2017enhancing} monitored students over nine semesters to see if reflexive weekly monitoring aids in completing course projects and improving coding skills. Ardimento et al.~\cite{ardimento2019evaluating} analyzed 40 novice developers' IDE usage patterns across five tasks. Romano et al.~\cite{romano2019empirical} examined 29 novice developers to determine the impact of test-driven development on positive affective states. Rehman et al.~\cite{rehman2020newcomer} reviewed 208 novice contributions on GitHub to identify common contribution types. Oliveira et al.~\cite{oliveira2020collaborative} compared novice and experienced developers to study code smells.
Compared with these studies, our work focuses on the software build process instead of general programming.

\section{Conclusion}
\label{sec:conclude}
We present a dual-phase exploratory study involving 55 CS students, which includes a resolution strategies phase and an intervention phase. The aim is to gain a comprehensive understanding of build issue resolution among CS students. We investigate the reported build issues and the strategies students use to resolve them. Due to students' lack of background knowledge and project-specific experience, their intuitive strategies tend to be straightforward. This limitation can make them more susceptible to certain build issues that are hard to resolve. Proactively preventing these potential \textit{trap issues} can significantly improve the success rate of their build outcomes, thereby enhancing the OSS learning experience in the SE course. This work highlights directions for further research in this area that could benefit various stakeholders in the OSS and CS education communities.

\IEEEtriggeratref{19}
\bibliographystyle{IEEEtran}
\bibliography{build_fail}

\begin{thebibliography}{10}
\providecommand{\url}[1]{#1}
\csname url@samestyle\endcsname
\providecommand{\newblock}{\relax}
\providecommand{\bibinfo}[2]{#2}
\providecommand{\BIBentrySTDinterwordspacing}{\spaceskip=0pt\relax}
\providecommand{\BIBentryALTinterwordstretchfactor}{4}
\providecommand{\BIBentryALTinterwordspacing}{\spaceskip=\fontdimen2\font plus
\BIBentryALTinterwordstretchfactor\fontdimen3\font minus \fontdimen4\font\relax}
\providecommand{\BIBforeignlanguage}[2]{{%
\expandafter\ifx\csname l@#1\endcsname\relax
\typeout{** WARNING: IEEEtran.bst: No hyphenation pattern has been}%
\typeout{** loaded for the language `#1'. Using the pattern for}%
\typeout{** the default language instead.}%
\else
\language=\csname l@#1\endcsname
\fi
#2}}
\providecommand{\BIBdecl}{\relax}
\BIBdecl

\bibitem{Gokhale2012}
S.~S. Gokhale, T.~Smith, and R.~McCartney, ``Integrating open source software into software engineering curriculum: Challenges in selecting projects,'' in \emph{2012 First International Workshop on Software Engineering Education Based on Real-World Experiences (EduRex)}, 2012, pp. 9--12.

\bibitem{Kotwani2011}
G.~Kotwani and P.~Kalyani, ``Open source software (oss): Realistic implementation of oss in school education,'' 12 2011.

\bibitem{Pinto2017}
G.~H.~L. Pinto, F.~F. Filho, I.~Steinmacher, and M.~A. Gerosa, ``Training software engineers using open-source software: The professors' perspective,'' in \emph{2017 IEEE 30th Conference on Software Engineering Education and Training (CSEE\&T)}, 2017, pp. 117--121.

\bibitem{nascimento2015open}
D.~M. Nascimento, R.~Almeida~Bittencourt, and C.~Chavez, ``Open source projects in software engineering education: a mapping study,'' \emph{Computer Science Education}, vol.~25, no.~1, pp. 67--114, 2015.

\bibitem{choi2021open}
E.~Choi, L.~Meng, and J.~Hott, ``Open source software practices in cs2,'' in \emph{Proceedings of the 21st Koli Calling International Conference on Computing Education Research}, 2021, pp. 1--5.

\bibitem{Salerno_de2023}
\BIBentryALTinterwordspacing
L.~Salerno, S.~de~França~Tonhão, I.~Steinmacher, and C.~Treude, ``Barriers and self-efficacy: A large-scale study on the impact of oss courses on student perceptions,'' in \emph{Proceedings of the 2023 Conference on Innovation and Technology in Computer Science Education V. 1}, ser. ITiCSE 2023.\hskip 1em plus 0.5em minus 0.4em\relax New York, NY, USA: Association for Computing Machinery, Jun. 2023, p. 320–326. [Online]. Available: \url{https://dl.acm.org/doi/10.1145/3587102.3588789}
\BIBentrySTDinterwordspacing

\bibitem{Fang_Endres_Zimmermann_Ford_Weimer_Leach_Huang_2023}
\BIBentryALTinterwordspacing
Z.~Fang, M.~Endres, T.~Zimmermann, D.~Ford, W.~Weimer, K.~Leach, and Y.~Huang, ``A four-year study of student contributions to oss vs. oss4sg with a lightweight intervention,'' in \emph{Proceedings of the 31st ACM Joint European Software Engineering Conference and Symposium on the Foundations of Software Engineering}, ser. ESEC/FSE 2023.\hskip 1em plus 0.5em minus 0.4em\relax New York, NY, USA: Association for Computing Machinery, Nov. 2023, p. 3–15. [Online]. Available: \url{https://dl.acm.org/doi/10.1145/3611643.3616250}
\BIBentrySTDinterwordspacing

\bibitem{Kwan_Schroter_Damian_2011}
I.~Kwan, A.~Schroter, and D.~Damian, ``Does socio-technical congruence have an effect on software build success? a study of coordination in a software project,'' \emph{IEEE Transactions on Software Engineering}, vol.~37, no.~3, p. 307–324, May 2011.

\bibitem{downs2012ambient}
J.~Downs, B.~Plimmer, and J.~G. Hosking, ``Ambient awareness of build status in collocated software teams,'' in \emph{2012 34th International Conference on Software Engineering (ICSE)}.\hskip 1em plus 0.5em minus 0.4em\relax IEEE, 2012, pp. 507--517.

\bibitem{phillips2014understanding}
S.~Phillips, T.~Zimmermann, and C.~Bird, ``Understanding and improving software build teams,'' in \emph{Proceedings of the 36th international conference on software engineering}, 2014, pp. 735--744.

\bibitem{kerzazi2014automated}
N.~Kerzazi, F.~Khomh, and B.~Adams, ``Why do automated builds break? an empirical study,'' in \emph{2014 IEEE International Conference on Software Maintenance and Evolution}.\hskip 1em plus 0.5em minus 0.4em\relax IEEE, 2014, pp. 41--50.

\bibitem{hilton2016continuous}
M.~Hilton, N.~Nelson, D.~Dig, T.~Tunnell, D.~Marinov \emph{et~al.}, ``Continuous integration (ci) needs and wishes for developers of proprietary code.(2016),'' 2016.

\bibitem{vassallo2020every}
C.~Vassallo, S.~Proksch, T.~Zemp, and H.~C. Gall, ``Every build you break: developer-oriented assistance for build failure resolution,'' \emph{Empirical Software Engineering}, vol.~25, pp. 2218--2257, 2020.

\bibitem{mcintosh2011empirical}
S.~McIntosh, B.~Adams, T.~H. Nguyen, Y.~Kamei, and A.~E. Hassan, ``An empirical study of build maintenance effort,'' in \emph{Proceedings of the 33rd international conference on software engineering}, 2011, pp. 141--150.

\bibitem{mcintosh2015large}
S.~McIntosh, M.~Nagappan, B.~Adams, A.~Mockus, and A.~E. Hassan, ``A large-scale empirical study of the relationship between build technology and build maintenance,'' \emph{Empirical Software Engineering}, vol.~20, pp. 1587--1633, 2015.

\bibitem{xia2014empirical}
X.~Xia, X.~Zhou, D.~Lo, X.~Zhao, and Y.~Wang, ``An empirical study of bugs in software build system,'' \emph{IEICE TRANSACTIONS on Information and Systems}, vol.~97, no.~7, pp. 1769--1780, 2014.

\bibitem{barrak2021builds}
A.~Barrak, E.~E. Eghan, B.~Adams, and F.~Khomh, ``Why do builds fail?—a conceptual replication study,'' \emph{Journal of Systems and Software}, vol. 177, p. 110939, 2021.

\bibitem{wu2020empirical}
Y.~Wu, Y.~Zhang, T.~Wang, and H.~Wang, ``An empirical study of build failures in the docker context,'' in \emph{Proceedings of the 17th international conference on mining software repositories}, 2020, pp. 76--80.

\bibitem{zhao2014empirical}
X.~Zhao, X.~Xia, P.~S. Kochhar, D.~Lo, and S.~Li, ``An empirical study of bugs in build process,'' in \emph{Proceedings of the 29th Annual ACM Symposium on Applied Computing}, 2014, pp. 1187--1189.

\bibitem{lou2020understanding}
Y.~Lou, Z.~Chen, Y.~Cao, D.~Hao, and L.~Zhang, ``Understanding build issue resolution in practice: symptoms and fix patterns,'' in \emph{Proceedings of the 28th ACM Joint Meeting on European Software Engineering Conference and Symposium on the Foundations of Software Engineering}, 2020, pp. 617--628.

\bibitem{huang2024build}
S.~Huang and X.~Wang, ``Build issue resolution from the perspective of non-contributors,'' in \emph{Proceedings of the 39th IEEE/ACM International Conference on Automated Software Engineering}, 2024, pp. 2304--2308.

\bibitem{script}
M.~Kerrisk, ``script - make typescript of terminal session,'' \url{https://man7.org/linux/man-pages/man1/script.1.html}, 2023.

\bibitem{Seaman_1999}
C.~Seaman, ``Qualitative methods in empirical studies of software engineering,'' \emph{IEEE Transactions on Software Engineering}, vol.~25, no.~4, p. 557–572, Jul. 1999.

\bibitem{github}
\BIBentryALTinterwordspacing
github, ``Github,'' 2022. [Online]. Available: \url{https://github.com/}
\BIBentrySTDinterwordspacing

\bibitem{openai2024chatgpt}
\BIBentryALTinterwordspacing
OpenAI, ``Chatgpt ({Feb 24 version}) [large language model],'' 2024. [Online]. Available: \url{https://chat.openai.com}
\BIBentrySTDinterwordspacing

\bibitem{kwan2011does}
I.~Kwan, A.~Schroter, and D.~Damian, ``Does socio-technical congruence have an effect on software build success? a study of coordination in a software project,'' \emph{IEEE Transactions on Software Engineering}, vol.~37, no.~3, pp. 307--324, 2011.

\bibitem{beller2017travistorrent}
M.~Beller, G.~Gousios, and A.~Zaidman, ``Travistorrent: Synthesizing travis ci and github for full-stack research on continuous integration,'' in \emph{2017 IEEE/ACM 14th International Conference on Mining Software Repositories (MSR)}.\hskip 1em plus 0.5em minus 0.4em\relax IEEE, 2017, pp. 447--450.

\bibitem{shridhar2014qualitative}
M.~Shridhar, B.~Adams, and F.~Khomh, ``A qualitative analysis of software build system changes and build ownership styles,'' in \emph{Proceedings of the 8th ACM/IEEE international symposium on empirical software engineering and measurement}, 2014, pp. 1--10.

\bibitem{zolfagharinia2017not}
M.~Zolfagharinia, B.~Adams, and Y.-G. Gu{\'e}h{\'e}nuc, ``Do not trust build results at face value-an empirical study of 30 million cpan builds,'' in \emph{2017 IEEE/ACM 14th International Conference on Mining Software Repositories (MSR)}.\hskip 1em plus 0.5em minus 0.4em\relax IEEE, 2017, pp. 312--322.

\bibitem{licker2019detecting}
N.~Licker and A.~Rice, ``Detecting incorrect build rules,'' in \emph{2019 IEEE/ACM 41st International Conference on Software Engineering (ICSE)}.\hskip 1em plus 0.5em minus 0.4em\relax IEEE, 2019, pp. 1234--1244.

\bibitem{lahtinen2005study}
E.~Lahtinen, K.~Ala-Mutka, and H.-M. J{\"a}rvinen, ``A study of the difficulties of novice programmers,'' \emph{Acm sigcse bulletin}, vol.~37, no.~3, pp. 14--18, 2005.

\bibitem{warner2017codepilot}
J.~Warner and P.~J. Guo, ``Codepilot: Scaffolding end-to-end collaborative software development for novice programmers,'' in \emph{Proceedings of the 2017 CHI Conference on Human Factors in Computing Systems}, 2017, pp. 1136--1141.

\bibitem{marques2017enhancing}
M.~Marques, S.~F. Ochoa, M.~C. Bastarrica, and F.~J. Gutierrez, ``Enhancing the student learning experience in software engineering project courses,'' \emph{IEEE Transactions on Education}, vol.~61, no.~1, pp. 63--73, 2017.

\bibitem{ardimento2019evaluating}
P.~Ardimento, M.~L. Bernardi, M.~Cimitile, and F.~M. Maggi, ``Evaluating coding behavior in software development processes: A process mining approach,'' in \emph{2019 IEEE/ACM International Conference on Software and System Processes (ICSSP)}.\hskip 1em plus 0.5em minus 0.4em\relax IEEE, 2019, pp. 84--93.

\bibitem{romano2019empirical}
S.~Romano, D.~Fucci, M.~T. Baldassarre, D.~Caivano, and G.~Scanniello, ``An empirical assessment on affective reactions of novice developers when applying test-driven development,'' in \emph{Product-Focused Software Process Improvement: 20th International Conference, PROFES 2019, Barcelona, Spain, November 27--29, 2019, Proceedings 20}.\hskip 1em plus 0.5em minus 0.4em\relax Springer, 2019, pp. 3--19.

\bibitem{rehman2020newcomer}
I.~Rehman, D.~Wang, R.~G. Kula, T.~Ishio, and K.~Matsumoto, ``Newcomer candidate: Characterizing contributions of a novice developer to github,'' in \emph{2020 IEEE international conference on software maintenance and evolution (ICSME)}.\hskip 1em plus 0.5em minus 0.4em\relax IEEE, 2020, pp. 855--855.

\bibitem{oliveira2020collaborative}
R.~Oliveira, R.~de~Mello, E.~Fernandes, A.~Garcia, and C.~Lucena, ``Collaborative or individual identification of code smells? on the effectiveness of novice and professional developers,'' \emph{Information and Software Technology}, vol. 120, p. 106242, 2020.

\end{thebibliography}
\end{document}